\documentclass[letterpaper, 10 pt, conference]{ieeeconf}  
\IEEEoverridecommandlockouts                              
\usepackage{newtxtext,newtxmath}
\usepackage{graphicx}
\graphicspath{{imm/}}
\usepackage{multirow}
\usepackage{amsmath}
\usepackage{acronym}

\title{Joint Distribution and Transitions of Pain and Activity in Critically Ill Patients}
\author{Florenc Demrozi, Graziano Pravadelli, Patrick J Tighe, Azra Bihorac and Parisa Rashidi
\thanks{F. Demrozi and G. Pravadelli are with the Computer Science Department, Univeristy of Verona, Italy
        {\small (name.surname@univr.it)}}
\thanks{P.J. Tighe is with the Department of Anesthesiology, College of Medicine, University of Florida, Gainesville, FL, USA {\small (ptighe@anest.ufl.edu)}}
\thanks{A. Bihorac is with the Department of Nephrology, College of Medicine, University of Florida, Gainesville, FL, USA        {\small (azra.bihorac@medicine.ufl.edu)}}
\thanks{P. Rashidi is with the Department of Biomedical Engineering, University of Florida, Gainesville, FL, USA
{\small (parisa.rashidi@ufl.edu)}}}
\usepackage{draftwatermark}
\SetWatermarkText{Accepted for Publication in EMBC 2020}
\SetWatermarkScale{0.3}
\begin{document}
\acrodef{DVPRS}{Defense and Veterans Pain Rating Scale}
\acrodef{ICU}{Intensive Care Unit}
\acrodef{STM}{State Transition Matrix}
\acrodef{SDM}{State Distribution Matrix}

\maketitle

\begin{abstract}
Pain and physical function are both essential indices of recovery in critically ill patients in the Intensive Care Units (ICU). Simultaneous monitoring of pain intensity and patient activity can be important for determining which analgesic interventions can optimize mobility and function, while minimizing opioid harm. Nonetheless, so far, our knowledge of the relation between pain and activity has been limited to manual and sporadic activity assessments. In recent years, wearable devices equipped with 3-axis accelerometers have been used in many domains to provide a continuous and automated measure of mobility and physical activity. In this study, we collected activity intensity data from 57 ICU patients, using the Actigraph GT3X device. We also collected relevant clinical information, including nurse assessments of pain intensity, recorded every 1-4 hours. Our results show the joint distribution and state transition of joint activity and pain states in critically ill patients.
\end{abstract}

\section{Introduction}\label{sec:intro}
After an Intensive Care Unit (ICU) stay, almost 50\% of patients described the perceived level of pain, at rest and during commonly performed procedures, between moderate and severe~\cite{chanques2007prospective,stanik2001pain,puntillo2002practices}. Typically, opioids and other pain-relief medications are prescribed to alleviate their pain and to maximize patient comfort. Nonetheless, a unilateral approach based on minimizing pain is increasingly seen as an insufficient for optimizing long-term patient outcomes. For example, early mobility has been shown to be associated with improved patient outcomes such as shorter length of stay.  Existing studies that examine the relation between pain and activity, typically rely on nurse-assessed activity scales, e.g. the PUMP-PLUS scale or other mobility assessment scales~\cite{titsworth2012effect,malheiro2015infectious}. Such information, while helpful, are assessed sporadically and can be subjective, while not taking into account activity information in between nurse visits. Furthermore, mobility assessment is not part of routine clinical care in all ICUs, further complicating investigation of activity in critically ill patients. In recent years, wearable devices equipped with accelerometer sensors have been used in many domains to measure activity intensity in a more objective and continuous manner~\cite{tipping2016icu,trost2011comparison,linton1985relationship}. Several previous studies have used accelerometers in the ICU to assess activity, however they have not been used for examining the relation between pain and activity. In this prospective observational study, we examined the relation between granular activity data and pain for the first time. We collected activity information of 57 patients using the Actigraph GTX3X device. Pain and other relevant clinical information were also collected from Electronic Medical Record (EMR) data.
\section{Background}\label{sec:back}
After an ICU stay, almost 50\% of patients described the perceived level of pain, at rest and during commonly performed procedures, between moderate and severe~\cite{chanques2007prospective,stanik2001pain,puntillo2002practices}. Currently, clinical pain decisions are generally oriented towards minimizing pain intensity. An overly narrow emphasis on pain intensity can lead to poor functional outcomes and may contribute to the ongoing opioid crisis~\cite{brummett2017new}. Recent efforts including ERAS (Enhanced Recovery After Surgery) and function-oriented pain assessments highlight this need by moving beyond simple indices of pain intensity~\cite{georgiou2015impact}. Simultaneous monitoring of pain intensity and function will allow clinicians to make decisions to optimize function that is impaired by pain, rather than just making decisions based on pain intensity alone. A common sensor for recording motor activities is the triaxial accelerometer. A triaxial accelerometer is a sensor that estimates the acceleration along the x, y, and z axes and from which velocity and displacement can also be determined. Accelerometers, nowadays, are used as motion detectors, body-position, and posture sensing, and generally in the human activity recognition context~\cite{ravi2005activity,wang2019deep}
\section{Methodology}\label{sec:meth}
In this study, we examined the distribution and state transition of joint pain and activity states in critically ill patients. We recruited 57 patients, and we measured their activity intensity using wearable Actigraph GTX3X device. The following sections provide details of subject recruitment, data preprocessing, and our association analysis.

\subsection{Subject Recruitment}\label{sec:sub_rec}
The subjects were recruited from the pool of critically ill patients admitted to the Shands Hospital at University of Florida. The study was approved under IRB 201400546. All patients provided written informed consent. Pain intensity was assessed using the Defense and Veterans Pain Rating Scale (DVPRS)~\cite{buckenmaier2013preliminary}, where each patient was asked to assess their pain intensity using a 0-10 scale. DVPRS was used as part of standard clinical practice and provides additional visual and contextual cues related to pain intensity. In the ICU, DVPRS is typically administrated every hour as part of clinical routine care. The frequency of pain assessment can change depending on patient’s condition. For example, sleeping patients might not be disturbed, or if patient has received medication to manage severe pain, pain might be assessed more frequently to observe also the medication effect. Finally, relevant clinical information such as length of stay were included.

\begin{table}[!t]
\centering
\caption{\small{Example of motor activity data captured from a tri-axial accelerometer.}}\label{tab:one}
\resizebox{0.45\textwidth}{!}{
\begin{tabular}{|c|c|c|c|c|c|c|}
\hline
Subject&Date&Time&Axis x&Axis y&Axis z&Magnitude\\
ID     &    & &$A_x$ &$A_y$ &$A_z$&$V_m$\\
\hline
31& 02/29/2017 &14:31:23.100 & 0  & 0 & 0&0   \\
31& 02/29/2017 &14:32:23.100 & 10 & 23&42&49  \\
31& 02/29/2017 &14:33:23.100 & 70 & 59&82&123 \\
\hline
\hline
\end{tabular}}
\end{table}
We recruited 57 patients in the surgical ICU at the quaternary academic University of Florida Health Hospital. For each patient, activity intensity information was collected using the Actigraph GTX3X device, which is a lightweight device (< 40 g), equipped with a 3-axial accelerometer sensor. Up to three Actigraph GTX3X devices were placed on the wrist, arm, and ankle. At patient request, or depending on the underlying medical conditions, injuries, or discomfort, one or two Actigraph devices were removed. In this study, we examined data from wrist placement. We initially recruited 91 patients. Due to Actigraph removal or unexpected ICU transfers, the final dataset included 57 patients who had complete wrist data.  Table~\ref{tab:one} shows data representation of the activity information. Columns one to three show the participant ID, date ($mm/dd/year$) and time ($hh:min:sec.ms$). Columns four to six show the activity count values ($A_x, A_y, A_z$) for the x, y, and z-axes of the Actigraph. Finally, column seven shows the vector magnitude obtained by applying Equation (1).
\begin{equation}
V_m=\sqrt{A_x^2+A_y^2+A_z^2}    
\end{equation}
\begin{table}[!bh]
\caption{\small{The Defense and Veterans Pain Rating Scale (DVPRS) scale and the discrete data binning intervals. $P_n$ refers to a specific level for the given binned schema.}}\label{tab:two}
\centering
\resizebox{0.45\textwidth}{!}{
\begin{tabular}{|c|c|c|c|c|c|}
\hline
Level&Descriptive&(1) Binning&(2) Binning &(3) Binning &(4) Binning\\
     & Category   &Schema $P_A$ &Schema $P_B$ & Schema $P_C$ &Schema $P_D$ \\\hline\hline
0 &No Pain &$P_1$ &$P_1$ &$P_1$ &$P_1$ \\ \hline
1 &Mild    &$P_1$ &$P_1$ &$P_2$ &$P_2$ \\ \hline
2 &Mild    &$P_1$ &$P_1$ &$P_2$ &$P_2$ \\ \hline
3 &Mild    &$P_1$ &$P_1$ &$P_2$ &$P_3$ \\ \hline
4 &Mild    &$P_1$ &$P_1$ &$P_2$ &$P_3$ \\ \hline
5 &Moderate&$P_2$ &$P_2$ &$P_3$ &$P_4$ \\ \hline
6 &Moderate&$P_2$ &$P_2$ &$P_3$ &$P_4$ \\ \hline
7 &Severe  &$P_2$ &$P_3$ &$P_4$ &$P_5$ \\ \hline
8 &Severe  &$P_2$ &$P_3$ &$P_4$ &$P_5$ \\ \hline  
9 &Severe  &$P_2$ &$P_3$ &$P_4$ &$P_6$ \\ \hline  
10&Severe  &$P_2$ &$P_3$ &$P_4$ &$P_6$ \\\hline\hline
\multicolumn{2}{|c|}{Nr. Categories} & 2&3&4&6\\ \hline\hline
\end{tabular}}
\end{table}

\subsection{Data Preprocessing}
To prepare our dataset for analysis, we performed several preprocessing tasks. The first step included removing incomplete observation windows, in which either pain or activity information are missing. In this case, this step will dismiss activity data that is not within an hour of any pain assessment.  At the end of this processing step, only intervals containing joint pain and activity information will be preserved (± 1 hour). This will allow us to construct joint pain and activity states required for our analysis.

\subsubsection{Discrete Data Binning}
In the literature, numerical pain scores are sometimes converted to the binned descriptive scores for ease of analysis (e.g. 0-2 = no pain, 3-6 = moderate pain, and 7-10 = severe pain)~c\cite{prkachin2008structure}. This will facilitate understanding the transitions between different activity/pain states.
For example, this will allow as to determine if most patients transition from a $\langle low activity, severe pain\rangle$ state (non-ideal) to $\langle high activity, mild pain\rangle$ state (ideal). Such information could have implications in terms of understanding patient pain trajectory and could be used for adjusting the physical therapy and other interventions.  In this study, we applied four discrete binning approaches for pain intensity to perform sensitivity analysis of our methodology (Table~\ref{tab:two}). The same discrete data binning approach was performed on activity data. Since there is not a well-defined activity threshold for critically ill patients, we took a data-driven approach.
\begin{table}[h!]
\caption{\small{Discrete data binning of activity levels. For sensitivity analysis, we will use two different binning schemas.}}\label{tab:three}
\centering
\resizebox{0.35\textwidth}{!}{
\begin{tabular}{|c|c|c|c|}
\hline
\multirow{3}{1.5cm}{Binning Schema (1)}
&Activity& Category $A_E$&Description\\\cline{2-4} 
&$Value \leqslant Median$& $A_1$&Low\\ 
&$Value > Median$        & $A_2$&High\\
\hline
\hline 
\multirow{4}{1.5cm}{Binning Schema (2)}
&Activity& Category $A_F$&Description\\\cline{2-4} 
&$Value \leqslant Q_1$   & $A_1$&Low\\
&$Q_1 < Value \leqslant Q_3$& $A_2$&Average\\
&$Value>Q_3$             & $A_3$&High\\
\hline
\hline
\end{tabular}}
\end{table}
\begin{figure}[!b]
\centering
\includegraphics[width=0.8\linewidth,page={1}]{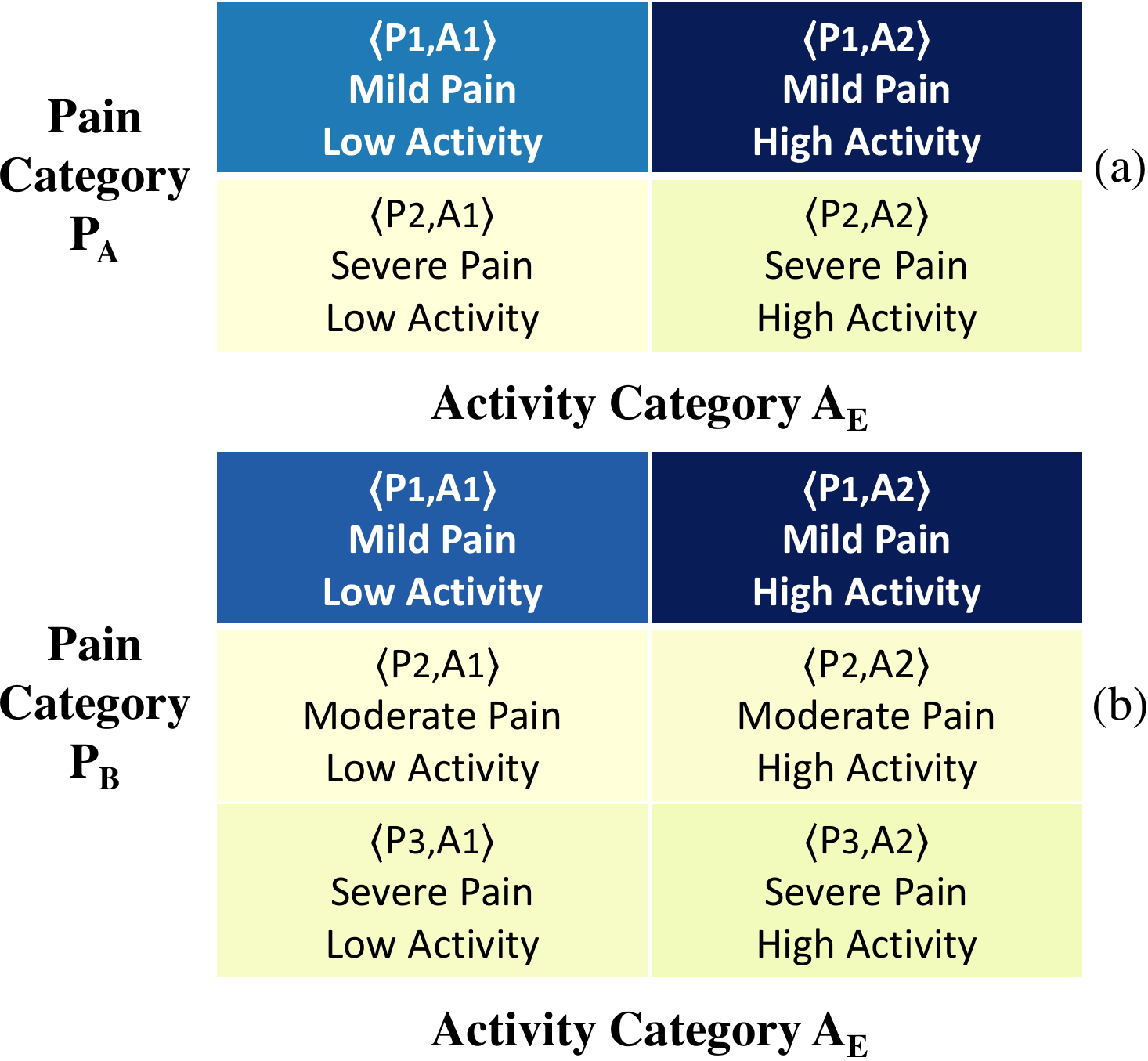}
\caption{Example State Distribution Matrix (SDM) shows the relative distribution of $\langle Pain , Activity\rangle$ states.}\label{fig:one}
\end{figure}
Table~\ref{tab:two} shows our binning schemas based on (a) Median (b) first quartile ($Q_1$), and (c) third quartile ($Q_3$), resulting in:
 \begin{itemize}
     \item $A_E$: two levels based on the median of activity vector magnitude value;
     \item $A_F$: three levels based on the first quartile ($Q_1$), median, and third ($Q_3$) quartile of activity vector magnitude value.
 \end{itemize}
 
\subsection{The Association between Pain and Activity}
\begin{figure}[!t]
\centering
\includegraphics[width=0.8\linewidth,page={2}]{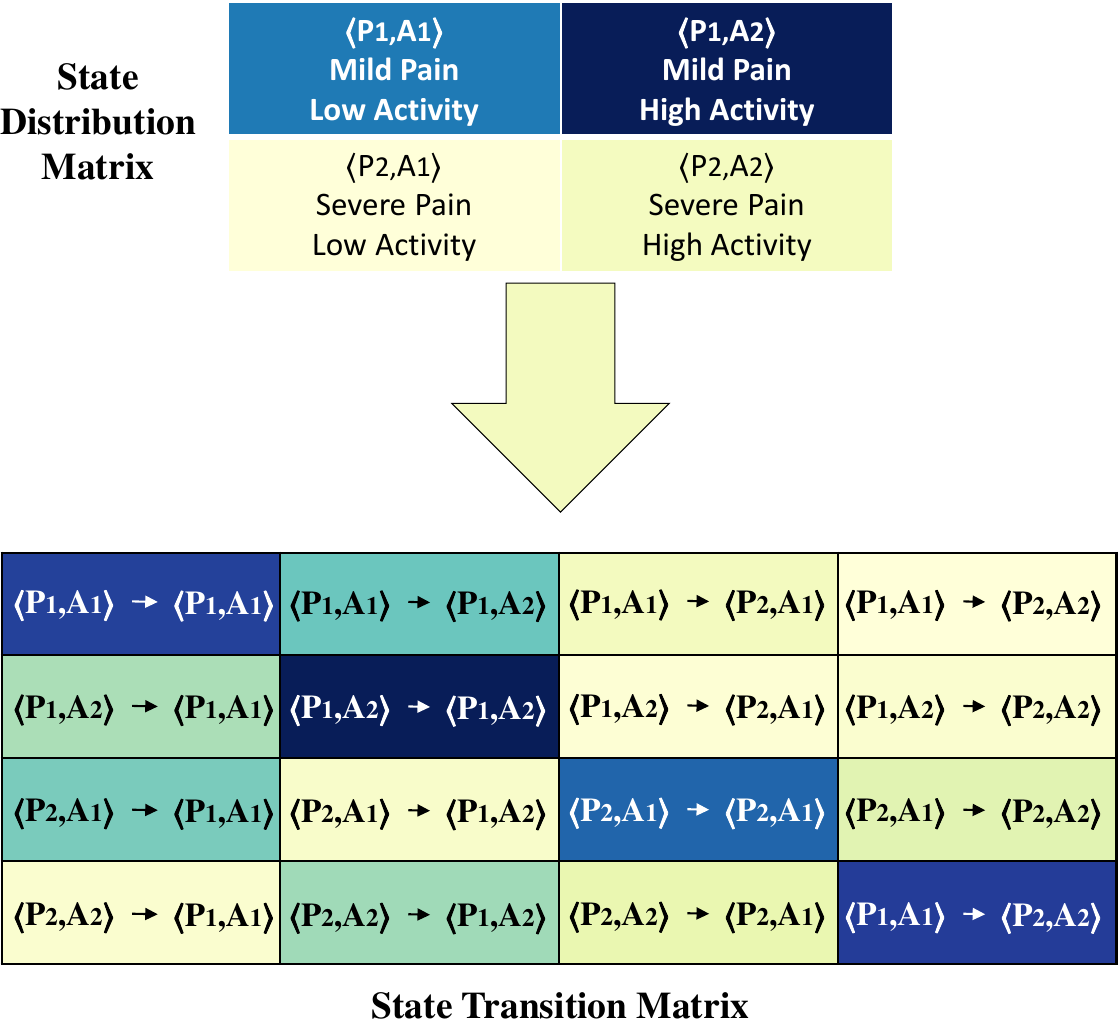}
\caption{Example State Transition Matrix (STM) shows the relative probability of transition from state $\langle P_i, A_j\rangle$ to state $\langle P_k, A_l\rangle$.}\label{fig:two}
\end{figure}

To examine the association between pain and activity, initially, we considered the relationship between 2-levels activity ($A_E$) and pain ($P_A$) data:
\begin{enumerate}
    \item State 1: $\langle Mild~Pain~,~Low~Activity\rangle $: Resting, sedation, reducing physical activity to avoid pain,
    \item State 2: $\langle Mild~Pain~,~High~Activity\rangle $: Functional, recovering,
    \item State 3: $\langle Severe~Pain~,~Low~Activity\rangle $: Critical,
    \item State 4: $\langle Severe~Pain~,~High~Activity\rangle $: Agitated, pain due to physical activity (e.g. physical therapy).
\end{enumerate}
The state pairs can be at higher granularity using discrete binning schema with more levels. For lack of space, in our analysis we will show only the association between between 2-levels activity ($A_E$) and pain ($P_A$) data. Examples of both matrices can be seen in Figure~\ref{fig:one} and Figure~\ref{fig:two}.\\
\textit{State Distribution Matrix (SDM)}. This matrix shows what percentage of time is spent in each of these states for all patients. In particular, for pain level $P_i$ and activity level $A_j$, this shows what percentage of time is spent in state $\langle P_i, A_j\rangle$, when normalized for all states.\\
\textit{State Transition Matrix (STM)}. This matrix shows the probability of transition from state $\langle P_i, A_j\rangle$ to state $\langle P_k, A_l\rangle$, when normalized for all transitions. Transition are formed when a new pain assessment is available. 
\section{Results}\label{sec:res}
This section will discuss the relationship between perceived pain and the amount of activity.
\begin{table}[!th]
\caption{\small{Cohort characteristics.}}\label{tab:four}
\centering
\begin{tabular}{|c|c|}
\hline
\multicolumn{2}{|c|}{All Participants (N=57)} \\\hline
Max Age              & 95                \\ \hline
Min Age              & 21                \\ \hline
Age(Std)             & 63.5 (17.3)       \\ \hline
Under 63 (\%)        & 21 (37\%)         \\ \hline
Over 63 (\%)         & 36 (63\%)         \\ \hline
Male (\%)            & 36 (63\%)         \\ \hline
Female (\%)          & 21 (37\%)         \\ \hline
Caucasian (\%)       & 51 (89\%)         \\ \hline 
Afro-American (\%)   & 4 (7\%)           \\ \hline
Other (\%)           & 2 (4\%)           \\ \hline
Hispanic (\%)        & 1 (2\%)           \\ \hline
Not-Hispanic (\%)   & 56 (98\%)         \\ \hline
\end{tabular}
\end{table}
\subsection{Dataset}
Starting from the dataset presented in Section III-A, since the patient’s data were sampled at different frequency, we up-sampled all the data to windows of 60 seconds (0.016 Hz). In total, the dataset presents approximately a total of 4000 hours of recorded data for a total of 240000 samples. Table~\ref{tab:four} shows cohort characteristics of recruited subjects, the majority of the patients (61\%) were enrolled in the study for less than four days and the time period between two consecutive pain assessments ranged from 10 minutes to 12 hours, with a median of 3 hours. Each pain sample is associated with the previous time window, of 1 hour, of activity data. Furthermore, the discharge outcomes are categorized as discharged to home and still hospitalized. 

In particular, in 24 (42\%) cases, the subjects were discharged from the hospital and sent home or to home care. In the remaining 33 (58\%) cases, the subjects continued hospitalization in other wards or hospitals. Throughout the dataset, there are time windows of variable length presenting zero motor activity . On average, 5.8\% of the data could correspond to time windows during which the subject did not wear the device. The non-wear time was considered as the time interval >2 hours with zero activity.

\subsection{Experimental Results}
We constructed the SDM and STM matrices with and without stratification according to gender, patient outcome (home discharge, other outcomes), and age ($\leqslant 63$, $> 63$).

\begin{figure}[!b]
\centering
\includegraphics[width=0.8\linewidth,page={3}]{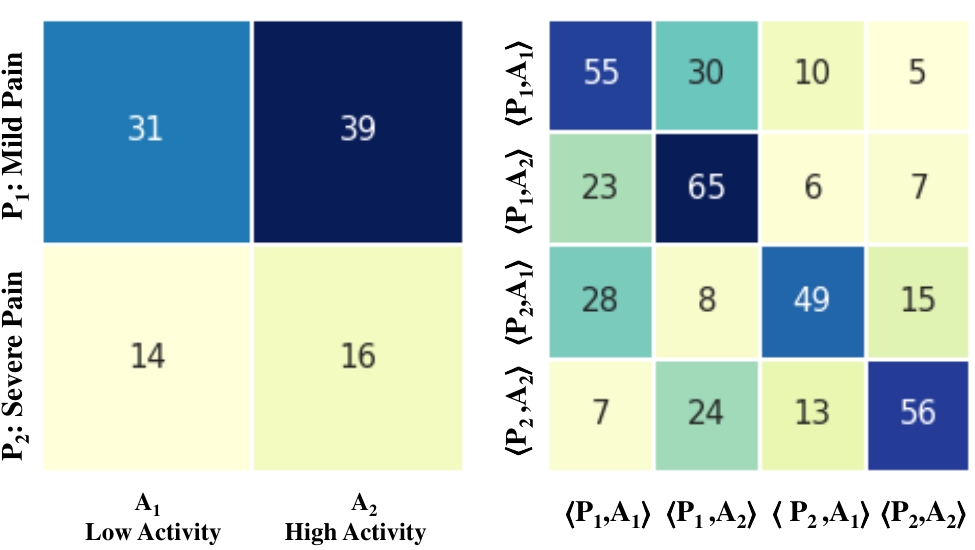}
\caption{Results for all patients without stratification, SDM and STM on two levels of pain and activity.}\label{fig:three}
\end{figure}
\subsubsection{No Stratification}
Figure~\ref{fig:three} and Figure~\ref{fig:four} shows the resulting SDM and STM matrices without any stratification for all the patients. In particular, Figure~\ref{fig:three} presents the results for two levels of pain ($P_A$) and activity ($A_E$). Figure~\ref{fig:four} presents the results for three levels of pain ($P_B$) and activity ($A_E$). 31\% of subjects show an association between mild pain levels and low activity levels. 39\% of subjects show an association between mild pain levels and high activity levels. 14\% of subjects show an association between severe pain levels and low activity levels. About 16\% of subjects show an association between severe pain levels and high activity levels. Furthermore, based on results showed from the STM we make the following observation:
\begin{figure}[!t]
\centering
\includegraphics[width=0.8\linewidth,page={4}]{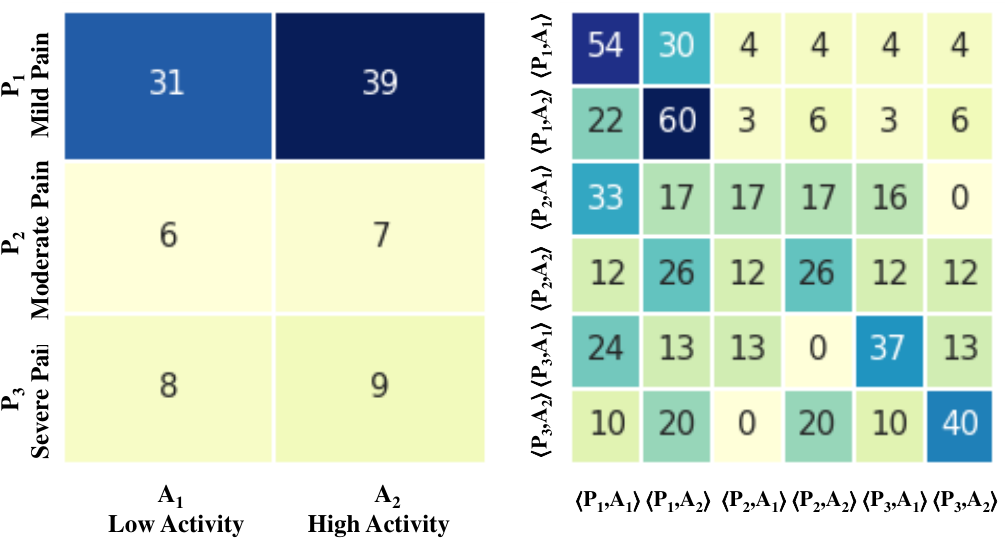}
\caption{Results for all patients without stratification, SDM and STM on three levels of pain and two levels of activity.}\label{fig:four}
\end{figure}
\begin{figure}[!b]
\centering
\includegraphics[width=0.8\linewidth,page={5}]{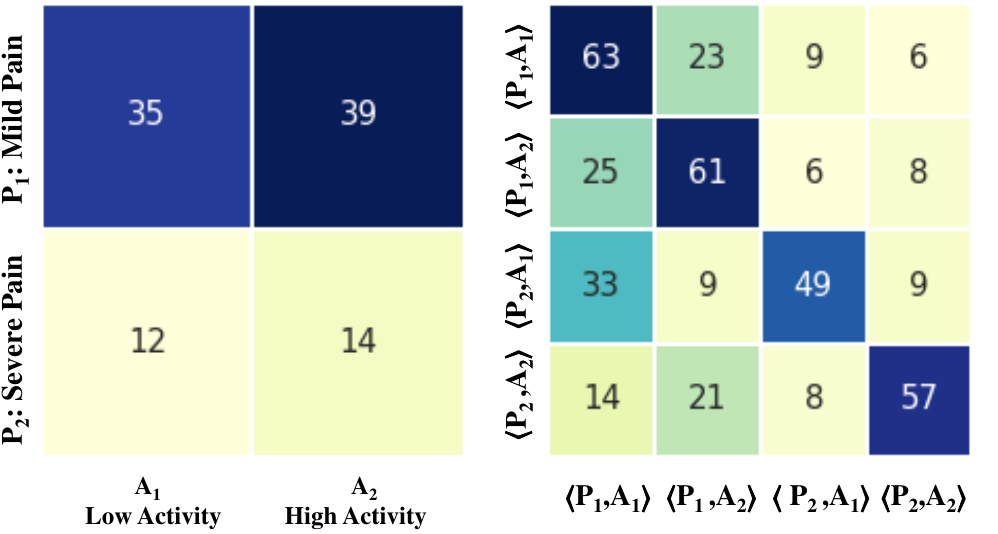}
\caption{SDM and STM on two levels of pain and activity differentiating by gender, 21 female patients.}\label{fig:five}
\end{figure}
\begin{itemize}
    \item The values on the diagonal are relatively larger than the values of the remaining cells in the respective rows. Thus, the probability of transitioning from one state to a different state is lower than the probability of remaining in the same state;
    \item The values on the lower triangular of the matrix show relatively higher values than those on the upper triangular. The probability of transitioning from a state of $\langle high activity, severe pain\rangle$ to a state of $\langle low activity, low pain\rangle$ is greater than the probability of transitioning from a $\langle low activity, low pain\rangle$ to a state of $\langle high activity, severe pain\rangle$, 7\% overall compared to 5\% once marginal probablity is computed over all such states.
    \item The probability of transitioning from a mild pain state to a severe pain state is lower than the probability of transitioning from a low activity state to a high activity state, 7\% overall compared to 14,5\%  once marginal probablity is computed over all such states;
    \item The probability of transitioning from a severe pain state to a mild pain state is higher than that of transitioning from a high activity state to a low activity state, 16,75\% overall compared to 12,25\% once marginal probablity is computed over all such states;
    \item The probability of transitioning from a severe pain state to a mild pain state is higher than that of transitioning from a mild pain state to a severe pain state, 16,75\% overall compared to 8,25\% once marginal probablity is computed over all such states.
\end{itemize}

\subsubsection{Gender Based Results}
\begin{figure}[!t]
\centering
\includegraphics[width=0.8\linewidth,page={6}]{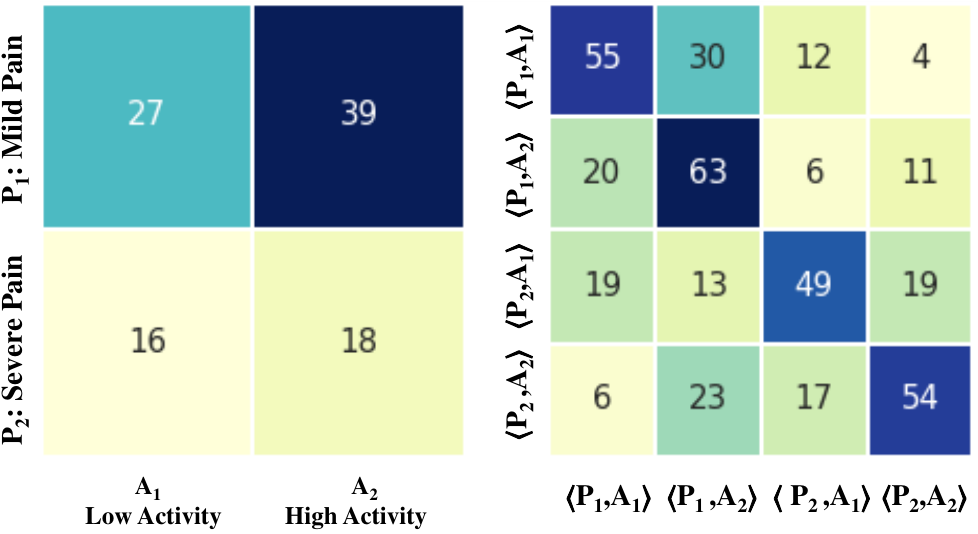}
\caption{SDM and STM on two levels of pain and activity differentiating by gender, 36 male patients.}\label{fig:six}
\end{figure}

Figure~\ref{fig:five} and Figure~\ref{fig:six} show the result when stratified by gender. Figure~\ref{fig:five} shows the results taking into account only 21 female subjects among the overall 57 subjects of the ICU dataset. Figure~\ref{fig:six} shows the results of taking into account the remaining 36 male subjects.  For both females and males, the SDM shows that the state distribution probability is almost the same. In particular, both low and severe pain is related to high activity. Furthermore, the STM upper triangle (Figure~\ref{fig:five}) related to the female group shows that the probability of transitioning from a mild pain state to a severe pain state is lower compared to the case of male subjects (Figure~\ref{fig:six}), 7,25\% overall compared to 8,25\%.
\begin{figure}[!bh]
\centering
\includegraphics[width=0.8\linewidth,page={7}]{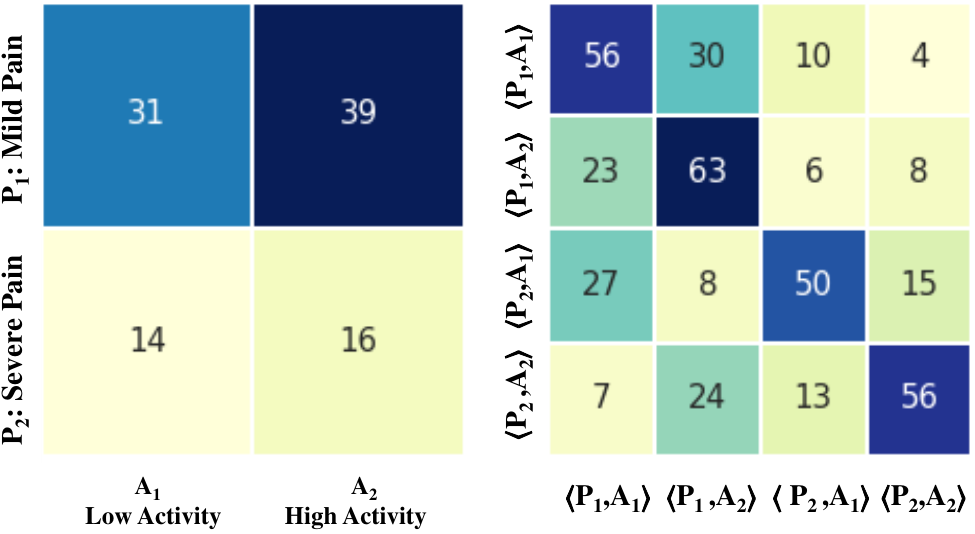}
\caption{Results obtained differentiating by outcome, 24 subjects discharged to home care.}\label{fig:seven}
\end{figure}
\subsubsection{Outcome and Age based results}
\begin{figure}[!t]
\centering
\includegraphics[width=0.8\linewidth,page={8}]{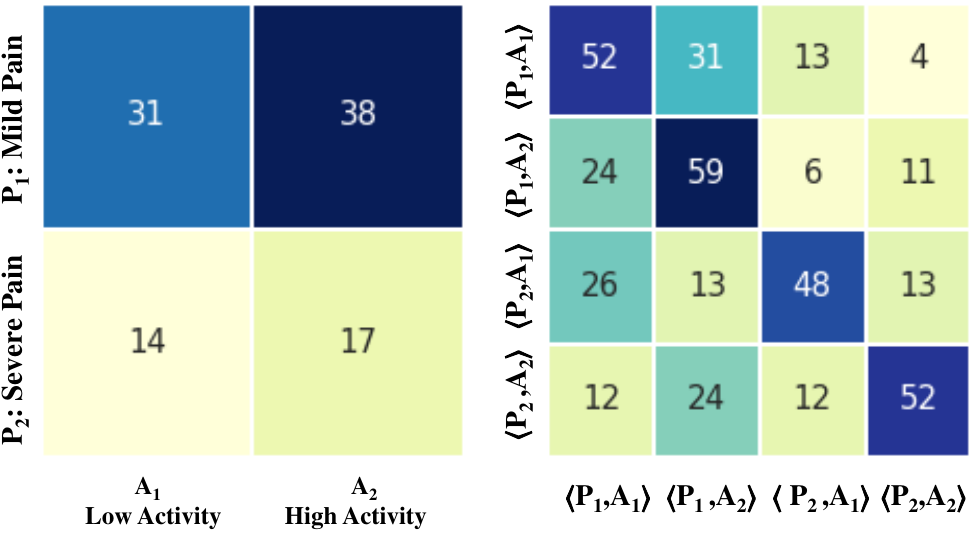}
\caption{Results obtained differentiating by outcome, 
33 subjects who continued hospital treatments.}\label{fig:eight}
\end{figure}

\begin{figure}[!t]
\centering
\includegraphics[width=0.8\linewidth,page={9}]{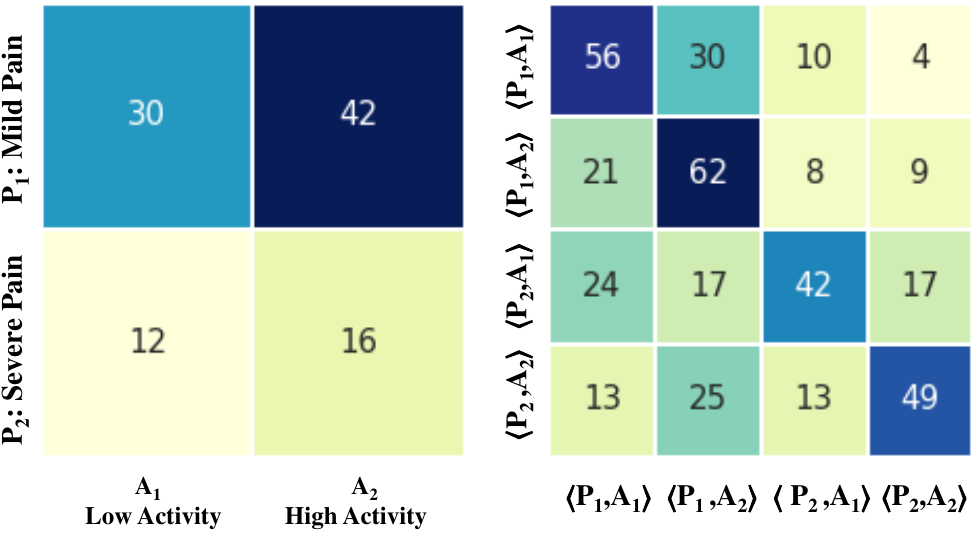}
\caption{Results obtained differentiating by age, 
21 patients older than 63 years.}\label{fig:nine}
\end{figure}

\begin{figure}[!t]
\centering
\includegraphics[width=0.8\linewidth,page={10}]{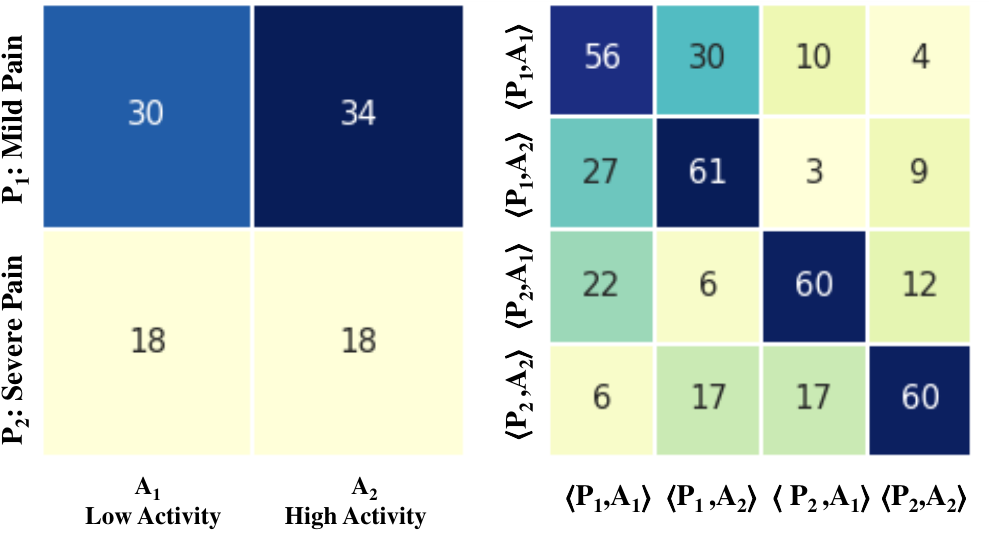}
\caption{Results obtained differentiating by age, 
36 patients younger than 63 years.}\label{fig:ten}
\end{figure}

Finally, Figure~\ref{fig:seven}, Figure~\ref{fig:eight}, Figure~\ref{fig:nine}, and Figure~\ref{fig:ten} show the SDM and STM matrices results stratified by discharge outcomes and age. Also, in these tests, we see that both high and mild pain levels are associated with a high activity level. This relationship (42\% / 16\%) is mainly observed in subjects over the age of 63. Furthermore, this group shows a lower probability of remaining in a state of severe pain (49\%) than that (60\%) shown by subjects under the age of 63. In the same way, we observe that the subjects discharged from the hospital, during the period of hospitalizations, show a higher probability of remaining in the same state of severe pain 56\%) than that shown by patients who remained hospitalized (52\%) even after data collection.

\section{Discussion}\label{sec:disc}
In this study, we examined the relation between sates formed from granular activity and pain data, and we also examined the transition amongst such states.  Our results show an association between mild pain levels and high activity levels for majority of patients (39\%). Nonetheless, the remaining patients exhibit either severe pain or low activity levels, which shows the need for joint optimization of pain management routines and early mobilization of patients.  The observed patterns deserve additional analyses and exploration to determine the nature of the observed correlations, including under the context of additional clinical and patient-oriented outcomes. In our future studies, we plan to examine these pain/activity state transitions in temporal manner, over the course of patient hospitalization.

\section{ACKNOWLEDGMENT}
AB and PR were supported by R01 GM110240 from the NIGMS, and by the National Institute of Biomedical Imaging and Bioengineering (grant R21EB027344-01). PR was supported by CAREER award, NSF-IIS 1750192, from the National Science Foundation (NSF), Division of Information and Intelligent Systems (IIS). PTJ and PR were supported by R01GM114290 from the NIGMS.
\bibliographystyle{IEEEtran}
\bibliography{IEEEabrv,biblio}{}

\end{document}